\documentclass[a4paper,12pt]{article}
\usepackage[left=2.5cm,bottom=3cm,right=2.5cm,top=3cm]{geometry} 

\usepackage{overpic,youngtab}
\usepackage{subfigure}
\usepackage[latin,english]{babel}
\usepackage{amsmath}
\usepackage{amssymb}
\usepackage{epstopdf}
\usepackage{graphics,psfrag,rotating}
\usepackage{mathabx}
\usepackage{graphicx}
\usepackage{dcolumn}
\usepackage{float}
\usepackage{pdflscape}
\usepackage{array}
\usepackage{booktabs}
\usepackage{amscd} 
\usepackage{mathtools}
\usepackage{fancybox}
\usepackage{fix-cm}
\usepackage[colorlinks=true,citecolor=blue,,linktocpage=true,linkcolor=blue,urlcolor=black]{hyperref}
\usepackage{tikz} 
\usetikzlibrary{matrix} 
\usetikzlibrary{positioning} 
\usetikzlibrary{arrows}
\usepackage{titlesec}
\usepackage{abstract}

\def\be{\begin{equation}}
\def\ee{\end{equation}}
\def\bea{\begin{eqnarray}}
\def\eea{\end{eqnarray}}
\def\pd{\partial}
\def\a{\alpha}
\def\b{\beta}

\def\g{\gamma}
\def\d{\delta}
\def\m{\mu}
\def\n{\nu}

\def\l{\lambda}

\def\r{\rho}

\def\s{\sigma}

\def\bi{\begin{itemize}}
\def\ei{\end{itemize}}

\newcommand{\email}[1]{\href{mailto:#1}{\tt #1}}

\begin{document}

		\vspace*{-1cm}
		\phantom{hep-ph/***} 
		{\flushleft
			{{FTUAM-19-6}}
			\hfill{{ IFT-UAM/CSIC-19-22}}}
		\vskip 1.5cm
		\begin{center}
		«{\LARGE\bfseries Conformal invariance versus  Weyl invariance }\\[3mm]
			\vskip .3cm
		
		\end{center}
		\vskip 0.5  cm
		\begin{center}
			{\large Enrique Alvarez, Jesus Anero  and Raquel Santos-Garcia }
			\\
			\vskip .7cm
			{
				Departamento de F\'isica Te\'orica and Instituto de F\'{\i}sica Te\'orica, 
				IFT-UAM/CSIC,\\
				Universidad Aut\'onoma de Madrid, Cantoblanco, 28049, Madrid, Spain\\
				\vskip .1cm

				\vskip .5cm
				\begin{minipage}[l]{.9\textwidth}
					\begin{center} 
						\textit{E-mail:} 
						\email{enrique.alvarez@uam.es}, \email{jesusanero@gmail.com}, \email{raquel.santosg@uam.es}
					
					\end{center}
				\end{minipage}
			}
		\end{center}
	\thispagestyle{empty}
	
\begin{abstract}\vspace{-1em}
	\noindent
The most general lagrangian describing spin 2 particles in flat spacetime and containing operators up to (mass) dimension 6 is carefully analyzed, determining the precise conditions for it to be invariant under linearized (transverse) diffeomorphisms, linearized Weyl rescalings, and conformal transformations.
\end{abstract}

\newpage
\tableofcontents
	\thispagestyle{empty}
\flushbottom

\newpage
	\setcounter{page}{1}
\section{Introduction}
Particle physics interactions, when considered at very high energy (probing then smaller and smaller distances), are expected to be independent of the individual masses of the particles themselves, which are negligible in comparison with the energy scale. Some sort of scale invariance is expected to be at work there. The same thing happens in second order phase transitions. The correlation length diverges, and again, scale invariance is at work. In fact in many known cases (in all unitary theories\footnote{There is a counterexample by Riva and Cardy \cite{Riva:2005gd} where scale invariance does not imply conformal invariance. The theory is the two-dimensional theory of elasticity which is not unitary.  }) this symmetry is upgraded to full conformal symmetry \cite{Zamolodchikov:1986gt}. There is however a caveat. In quantum field theory the well-known need to renormalize the bare quantities implies that an arbitary mass must be introduced. This is the origin of the dependence of coupling constants with the energy scale, encoded in the correspondiong beta-functions.
\par
It is however only recently that the precise relationship between scale invariance, conformal invariance and Weyl invariance has been clarified (cf. \cite{Dymarsky,Nakayama:2013is} and references therein).
This includes the precise conditions for  scale invariant theories to become  conformal invariant  and also  the existence of the so-called a-theorem for renormalizable theories (cf.\cite{Komargodski}and \cite{Shore:2016xor} for a recent review). Most of the work done so far has been in flat spacetime, where the gravitational field is absent, or at most, non-dynamic.
\par
 When such a gravitational field is present \cite{Alvarez} there are two possible generalizations of scale invariance. The most direct of those is the algebra of {\em conformal Killing vector fields} (CKV), that is, those that obey
 \be
 \mathcal{L}(\xi)g_{\m\n}=\phi(x) g_{\m\n}
 \ee
 the fact that
 \be
 \mathcal{L}(\left[\xi,\eta\right])=\mathcal{L}(\xi)\mathcal{L}(\eta)-\mathcal{L}(\eta)\mathcal{L}(\xi)
 \ee
 implies that the set of all CKV generate an algebra, which for Minkowski spacetime is the conformal algebra, $SO(2,n)$. In fact, the maximal possible dimension of the conformal algebra is precisely
 \be
 d={(n+1)(n+2)\over 2}
 \ee
 which is attained by conformally flat spacetimes (the ones with vanishing Weyl tensor \cite{Weyl}). Unfortunately, however, this property is not generic; that is, an arbitrary metric does not support any CKV, and the corresponding algebra has to be studied for each particular spacetime by itself.
 \par
 The next  most natural symmetry to study is Weyl invariance, the invariance of the action under local rescalings of the metric tensor. 
 \be
 g_{\m\n}(x)\rightarrow \Omega^2(x)\,g_{\m\n}(x)
 \ee
 this invariance, besides, can still be studied in the linear limit, when the gravitational field manifests itself as a perturbation of the Minkowski metric.
 \be
 g_{\m\n}(x)= \eta_{\m\n} + \kappa h_{\m\n}(x)
 \ee
Given that the previous expansion is exact, the linearized Weyl symmetry of the metric perturbation can be written as
\be
\kappa \,  \d h_{\m\n}= 2 \omega(x)\eta_{\m\n} + 2 \kappa \, \omega \, \mathcal{L}_{\xi} h_{\m\n}
\ee
Let us remark the appearance of an order $h$ piece in the variation of the perturbation that will be relevant in our analysis. 
\par
In order for a flat spacetime theory to be scale invariant, the  (Rosenfeld) energy-momentum tensor must be a total derivative (on-shell)
\be
\eta^{\m\n} T_{\m\n}=\pd_\m V^\m
\label{virial}
\ee
where $V^\m$ is the {\em virial current} \cite{Callan,Coleman, Osborn}. This is enough to guarantee the existence of a conserved scale  current
\be
j^\m\equiv x^\l T_\l^\m-V^\m
\ee
In the particular case when the virial current is itself a divergence, that is, when
\be
V^\m=\pd_\n \s^{\m\n}
\label{derivativevirial}
\ee 
then the theory is conformally invariant under the group $O(2,n)$, and the conserved current reads
\be
K^{\m\n}\equiv \left(2 x^\n x_\r-x^2 \d^\n_\r\right)T^{\r\m}-2 x^\n V^\m+2 \s^{\m\n}
\ee
which also implies that the energy-momentum tensor can be improved.
\par

\par

In the present paper we want to clarify the precise relationship between Weyl invariant theories  (WIFT) and conformal invariant theories (CFT) for systems  where the gravitational field is dynamic, but still approachable as a fluctuation of flat spacetime. Our analysis then concerns mostly spin 2 theories in flat spacetime as described by a rank two symmetric field in Minkowski space. Our plan is to do it systematically, determining the conditions for scale invariance (which is still meaningful in flat space), conformal invariance and Weyl invariance. 

We analyze first the most general lagrangian containing dimension 4 operators, and then we do the same analysis for dimension 6 operators, containing two and four derivatives respectively (operators appearing in the weak field expansion of gravitational theories linear and quadratic in the curvature). We then study dimension 5 and dimension 6 operators with two derivatives.  The analysis is, in some sense, the continuation of the one in \cite{ABGV, VanNiewu}. 
Recent works regarding conformal invariance and Weyl invariance include \cite{Karananas, Farnsworth:2017tbz, Wu:2017epd}.
We are always (with the only exception of our discussion of the improvement of the energy-momentum tensor) refering to actions defined as spacetime integrals, so that we allow for integration by parts, in other words, when we claim that an expression vanishes, we mean this only {\em up to total derivatives}.
\par

\section{Symmetries of the low energy spin 2 action}

Let us revisit the possible symmetries of low dimension  kinetic operators in spin 2 theories \cite{ABGV} in flat spacetime where the graviton is represented by a symmetric tensor $h_{\m\n}$. Lorentz invariance will be  assumed throughout the paper.
If we want the field equations to be given by differential operators of (at least) second order, then  the lagrangian  has got  to incorporate at least two derivatives. Let us begin our study with the operators of lowest possible dimension.
\subsection{Dimension 4 operators}
Our building blocks are the gravitational field, $h_{\a\b}$ (assumed to be of mass dimension 1) and the spacetime derivatives, $\pd_\m$. There are four different dimension four operators with two derivatives
\bea
&&{\cal D}_1\equiv{1\over 4} \pd_\m h_{\n\r} \pd^\m h^{\n\r}\nonumber\\
&&{\cal D}_2\equiv -{1\over 2} \pd^\l h_{\l\r}\pd_\s h^{\r\s}\nonumber\\
&&{\cal D}_3\equiv {1\over 2} \pd^\l h \pd^\s h_{\l\s}\nonumber\\
&&{\cal D}_4\equiv-{1\over 4}\pd_\m h \pd^\m h
\label{4dimop}
\eea
There is another operator
\be
{\cal D}_5\equiv  -{1\over 2}\pd_\s h_{\l\r}\pd^\l h^{\r\s}
\ee
which is equivalent to ${\cal D}_2$ modulo total derivatives:
\be
{\cal D}_5={\cal D}_2-{1\over 2}\pd_\l\left(\pd_\s h^\l_\r h^{\r\s}\right)+{1\over 2}\pd_\s\left(\pd_\l h^{\l\r} h_\r^\s\right)
\ee
Then, the most general action principle  involving dimension 4 derivative operators reads 
\be
S= \int d^4 x\,\sum_{i=1}^{4} \a_i\, {\cal D}_i
\ee

First, we are going to consider invariance under linearized diffeomorphisms, {\em LDiff} gauge symmetry. This is the one implemented in the pioneering work by Fierz-Pauli 
\be
\d h_{\m\n}=\pd_\m \xi_\n+\pd_\n\xi_\m
\ee
The variation of the fragment of the action containing the ${\cal D}$ terms only (that is, with $m_i^2=\l_\a=0$) and upon integration by parts, yields
\bea
&&\d \int d^4 x\,{\cal D}_1=\int d^4 x\,\xi^\l\,\Box\, \pd^\n\, h_{\n\l}\nonumber\\
&&\d \int d^4 x\,{\cal D}_2=-\int d^4 x\,\xi^\l\left(\pd_\l\,\pd^\r\pd^\s h_{\r\s}+ \Box\,\pd^\s h_{\l\s}\right)\nonumber\\
&&\d \int d^4 x\,{\cal D}_3=\int d^4 x\,\xi^\l\left(\pd_\l \pd^\r\pd^\s h_{\r\s}+\pd_\l\Box h\right)\nonumber\\
&&\d \int d^4 x\,{\cal D}_4=-\int d^4 x\,\xi^\l\, \pd_\l\, \Box h	
\eea
so that linearized diffeomorphisms ({\em LDiff} henceforth) imposes some relations among the coupling constants
\bea
\a_1=\a_2=\a_3=\a_4
\eea
this is, $\a_i=1\,\forall i$.

\par
The second case we are analyzing is the invariance under transverse linearized diffeomorphisms ({\em LTDiff} henceforth), that is, diffeomorphisms such that their generating vector fields obey
\be
\pd_\m \xi^\m=0
\ee
These conditions impose
\bea
\a_1=\a_2
\eea
but allow for arbitrary values of $\a_3$ and $\a_4$.
\par
In third place, under linearized Weyl transformations, {\em LWeyl}, the variation of the metric reads
\be  \d h_{\a\b}= \dfrac{2}{\kappa} \, \omega(x)\eta_{\a\b}
\ee
and after integration by parts\footnote{Although we do not write the integrals explicitely, integration by parts is carried out in the analysis and total derivatives are not considered as stated in the introduction.}, 
\bea
&&\d{\cal D}_1=-{1\over 2}\,\omega\, \Box h\nonumber\\
&&\d{\cal D}_2=\omega\,\pd^\a\pd^\b h_{\a\b}\nonumber\\
&&\d{\cal D}_3=- {\omega\over 2}\left(4 \pd^\a\pd^\b h_{\a\b}+\Box h\right)\nonumber\\
&&\d{\cal D}_4=2 \omega \,\Box h
\label{Weylvar1}
\eea
where we have multiplied by $\dfrac{\kappa}{2}$ for simplicity. 
The invariance under LWeyl puts further constraints on the coupling constants, namely
\bea
&&\a_1+\a_3-4 \a_4=0\nonumber\\
&& 4 \a_3- 2 \a_2=0\nonumber\\
\label{4dWeyl}
\eea
 In \cite{ABGV} we have dubbed {\em WTDiff} to the theory with {\em TDiff} invariance enhanced with linearized Weyl symmetry, {\em LWeyl}. This is the particular case of the above, corresponding to  $\a_1=\a_2=1$ and
\bea
&&\a_3=\dfrac{1}{ 2}, \quad \quad \a_4=\dfrac{3}{ 8}
\eea
A consistent  non-linear completion of the actions which fullfill these requirements are the ones explained in the appendix \eqref{WTDiff} namely actions proportional to
\be
S_{WTDiff} = - \dfrac{1}{ \kappa^2}\int d^4 x \, g^{1/4} \left(R + \dfrac{3}{32} \, \dfrac{(\nabla g)^2}{g^2}\right)
\label{wtdiff}
\ee

Finally, we could consider only traceless graviton fields, $h_{\a\b}$ such that $h\equiv \eta^{\a\b} h_{\a\b}=0$. Obviously, in this case, ${\cal D}_3={\cal D}_4=0$, and for consistency, we can only implement {\em TDiff} with the coupling constants fixed to
\be
\a_1=\a_2=1
\ee
\subsubsection{Scale and conformal invariance}
The most general lagrangian we are considering  (without the mass terms) is obviously {\em scale invariant} under
\bea
&&x^\m\rightarrow \l x^\m \nonumber\\
&&h_{\m\n}\rightarrow \l^{-1} h_{\m\n}
\eea
 with the assigned scaling dimensions. 
In order to make a full analysis of the scale and conformal invariance of the theory we have to compute the energy momentum tensor of these theories. In this case, and neglecting total derivatives, the metric (or Rosenfeld's) energy momentum tensor has the form 
\bea
T_{\m\n}&=& {1\over 4}\a_1\bigg\{\pd_\m h_{\a\b}\pd_\n h^{\a\b}+2\pd_\l h_{\m\b}\pd^\l h_\n\,^\b\bigg\}-{1\over 2}\a_2\bigg\{\pd_\m h_{\n\l}\pd_\s h^{\l\s}+\pd_\n h_{\m\l}\pd_\s h^{\l\s}+\pd^\l h_{\l\m}\pd^\d h_{\d\n}\bigg\}\nonumber\\
&&+{1\over 2}\a_3\bigg\{\pd^\a h_{\m\n}\pd^\l h_{\a\l}+{1\over 2}\left(\pd^\a h \pd_\m h_{\a\n}+\pd^\a h \pd_\n h_{\m\a}+\pd_\m h \pd^\l h_{\l\n}+\pd_\n h \pd^\l h_{\l\m}\right)\bigg\}\nonumber\\
&&-{1\over 4}\a_4\bigg\{2 \pd^\a h \pd_\a h_{\m\n}+\pd_\m h \pd_\n h\bigg\}-{1\over 2} L\eta_{\m\n}
\eea
whose trace reads
\be
T={3\over 4} \a_1 \,(\pd_\m h_{\a\b})^2-{3\over 2} \a_2\,\pd_\l h^{\l\s} \pd^\r h_{\r\s}+{3\over 2} \a_3 \pd^\a h \pd^\l h_{\l\a}-{3\over 4}\a_4\,(\pd_\m h)^2-2\,L=L\label{T}
\ee
The equations of motion (eom) read
\be
\dfrac{\d S}{ \d h^{\a\b}}=-{\a_1\over 2}\Box h_{\a\b}+{\a_2\over 2}\pd_\a\pd^\l h_{\l\b}+{\a_2\over 2}\pd_\b\pd^\l h_{\l\a}-{\a_3\over 2} \eta_{\a\b} \pd^\l\pd^\s h_{\l\s}-{\a_3\over 2} \pd_\a\pd_\b h+{\a_4\over 2} \eta_{\a\b} \Box h
\ee
\par
In fact we know that,  as explained in the introduction, scale invariance implies that the trace of the energy momentum tensor can be written, on-shell, as a total derivative, that is, as the divergence of the virial \eqref{virial}. What happens is that
 whereas two lagrangians that differ by a total derivative still generate the same eom, the specific form of the virial does depend on the particular form of the lagrangian. This, in turn, also determines whether the virial itself can be written as a total derivative. Examining the contributions of the whole set of total derivative operators, leads to the convenient action
\be
S= \int d^4 x\left[{\a_1\over 4} \pd_\m h_{\n\r} \pd^\m h^{\n\r}-{\a_2\over 4} \left(\pd^\l h_{\l\r}\pd_\s h^{\r\s}+\pd_\s h_{\l\r} \pd^\l  h^{\r\s}\right)+{\a_3\over 2} \pd^\l h \pd^\s h_{\l\s}-{\a_4\over 4}\pd_\m h \pd^\m h\right]
\ee
The trace of the energy momentum tensor can then be rewritten as
\bea
T &= & \left\lbrace \a_1\left[{1\over 4} \pd^\m [h^{\n\r}\pd_\m h_{\n\r}] -{1\over 4}  h_{\m\n} \Box h^{\m\n}\right]+ \right. \nonumber\\
&&+\a_2\left[-{1\over 4}\pd_\s[ h^{\r\s} \pd^\l h_{\l\r}]-{1\over 4}\pd_\s [h_{\l\r}  \pd^\l h^{\r\s}]+{1\over 4} h^{\r\s} \pd_\s\pd^\l h_{\l\r}+{1\over 4}  h_{\l\r}\pd^\l\pd_\s h^{\r\s}\right]+\nonumber\\
&&+\a_3\left[{1\over 4}\pd^\s [h_{\l\s} \pd^\l h] +{1\over 4} \pd^\l [h \pd^\s h_{\l\s}]-{1\over 4}  h_{\l\s}\pd^\s\pd^\l h -{1\over 4} h \pd^\l \pd^\s h_{\l\s}\right]+\nonumber\\
&&+ \left. \a_4\left[-{1\over 4}\pd_\m[ h \pd^\m h]+{1\over 4} h \Box h\right] \right\rbrace 
\eea
so that on-shell, we can express the trace as $T =\pd_\m V^\m$, where
\be
V^\m=\left\lbrace {\a_1\over 4}  h^{\n\r}\pd^\m h_{\n\r}-{\a_2\over 4}h^{\r\m} \pd^\l h_{\l\r}-{\a_2\over 4} h_{\l\r}\pd^\l h^{\r\m}+{\a_3\over 4}h_{\l}^\m \pd^\l h +{\a_3\over 4} h \pd^\s h^\m_{\s}-{\a_4\over 4}h \pd^\m h \right \rbrace 
\ee
\par
It is also well-known that a conformal current can be formally constructed 
in case the virial can be so expressed as $V^\m = \pd_\n \s^{\m\n}$. In our case, and upon using the eom, 
\be
\s^{\m\n}=\left\lbrace{\a_1\over 8}  \eta^{\m\n}h^{\a\b} h_{\a\b}-{\a_2\over 4}h^{\r\m}  h^\n_{\r}+{\a_3\over 4}h^{\m\n} h -{\a_4\over 8} \eta^{\m\n} h^2 \right\rbrace
\ee
In that way we get such a tensor  (on-shell), for any value of the coupling constants\footnote{
We insist that this result (because this tensor  is after all a total derivative) depends on the boundary terms that are neglected in order to write down the original lagrangian. The monomials have to be written down as indicated; to be specific, in order to split $\mathcal{D}_2$ in the two pieces, there is an integration by parts
\be
- \pd^\l h_{\l\r}\pd_\s h^{\r\s} - \pd^\l h_{\l\r}\pd_\s h^{\r\s} = - \pd^\l h_{\l\r}\pd_\s h^{\r\s} - \pd_\s h_{\l\r}\pd^\l h^{\r\s} - \pd^\l (h_{\l\r} \pd_\s h^{\r\s})  + \pd_\s (h_{\l\r} \pd^\l h^{\r\s}) 
\ee
so we have two total derivatives appearing in order to interchange the two derivatives. Although this does not contribute to the equations of motion, it does contribute to the virial, $L = L' + \pd_\m j^\m \rightarrow V^\m = V'^\m + j^\m$. It could well be the case that $V^\m = \pd_{\n} \s^{\m\n}$ but $V'^\m + j^\m \neq \pd_{\n} \s'^{\m\n}$. If we take this contributions into account one of the total derivatives cancels one of the pieces of the virial proportional to $\a_2$  and the other one is summed with the other piece. We end up with
\be
V^\m=  \left\lbrace {\a_1\over 4}  h^{\n\r}\pd^\m h_{\n\r}-{\a_2\over 2}h^{\r\m} \pd^\l h_{\l\r}+{\a_3\over 4}h_{\l}^\m \pd^\l h +{\a_3\over 4} h \pd^\s h^\m_{\s}-{\a_4\over 4}h \pd^\m h \right\rbrace\ee
and we could not write this as the derivative of a two-index tensor.}.
\par
Once a particular form of an $\s^{\m\n}$ is found, there is a systemaytic way of improving the energy-momentum tensor \cite{Callan}. The  improvement consists on adding another piece to the initial energy momentum tensor, so that the trace of the new energy-momentum tensor is precisely cero, that is, we avoid the total derivative terms. The piece in \cite{Callan} has the form
\be
\Theta^{\m\n} = T^{\m\n} + \dfrac{1}{2}\pd_{\l}\pd_{\r}X^{\l \r \mu \nu}
\ee
where $X^{\l \r \mu \nu}$ is symmetric $(\m,\n)$, and divergenceless. The precise for of the improvement reads
\be
X^{\l \r \mu \nu} = g^{\l\r} \s_+^{\m\n}- g^{\l\m} \s_+^{\r\n}- g^{\l\n} \s_+^{\m\r}+ g^{\m\n} \s_+^{\l\r}-\dfrac{1}{3} g^{\l\r}  g^{\m\n} \s_{+\a}^\a +  \dfrac{1}{3} g^{\l\m}  g^{\r\n} \s_{+\a}^\a
\ee
where $\s_+^{\m\n}$ stands for the symmetric part of $\s^{\m\n}$. The original analysis \cite{Coleman} was specific for  $n=4$ dimensions but it can be generalized to arbitrary dimension \cite{Polchinski:1987dy}.
\subsection{Dimension 6 operators}
In this section, we want to study the dimension 6 operators, and among these, there are various possibilities. First  we take operators with four derivatives and two $h_{\a\b}$. After integration by parts, those are
\bea\label{seis}
&{\cal O}_1\equiv h_{\a\b}\left(\pd^\a \pd^\b \pd^\g \pd^\d\right)h_{\g\d}\nonumber\\
&{\cal O}_2\equiv h_{\a\b}\left(\pd^\a\pd^\b \eta^{\g\d}\Box\right)h_{\g\d}\nonumber\\
&{\cal O}_3\equiv h_{\a\b}\left(\pd^\a \pd^\g \eta^{\b\d}\Box\right)h_{\g\d}\nonumber\\
&{\cal O}_4\equiv h_{\a\b}\left(\Box^2 \eta^{\a\g}\eta^{\b\d}\right)h_{\g\d}\nonumber\\
&{\cal O}_5\equiv h_{\a\b}\left(\Box^2\eta^{\a\b}\eta^{\g\d}\right)h_{\g\d}
\label{opquadratic}
\eea
There is a small caveat here. There are also many operators with two derivatives and four $h_{\a\b}$, that will be analyzed in the next section. It is the case that these operators do not appear as a limit of quadratic {\em Diff} or {\em TDiff} invariant theories; they can only appear as higher order contributions to lagrangians linear in the curvature. 

Let us consider the general theory involving dimension 6 operators which can come from quadratic theories of gravity, namely
\be
S_{quad}= \kappa^2 \int d^4 x \,\sum_{i=1}^5  g_i\, {\cal O}_i
\ee

\par
Like in the previous section,  we first study the  {\em LDiff} symmetry,  upon which the ${\cal O}$ transform as
\bea
&&\d {\cal O}_1=-4 \xi_\l\pd^\l\pd^\a\pd^\b\Box h_{\a\b}\nonumber\\
&&\d {\cal O}_2=-2 \xi_\l \left(\pd^\l \Box^2 h+ \pd^\l\pd^\a\pd^\b \Box h_{\a\b}\right) \nonumber\\
&&\d {\cal O}_3=-2\xi^\l\left(\pd^\b\Box^2 h_{\l\b}+\pd_\l\pd^\a\pd^\b\Box h_{\a\b}\right)\nonumber\\
&&\d {\cal O}_4=-4\xi^\l \pd^\a \Box^2 h_{\a\l}\nonumber\\
&&\d {\cal O}_5=-4 \xi^\l \pd_\l \Box^2 h
\eea
after having integrated by parts. Then the symmetry under $LDiff$ imposes the following relations between the coupling constants
\bea
&&2g_1+g_2+ g_3=0\nonumber\\
&&g_2+2 g_5=0\nonumber\\
&&g_3+ 2 g_4=0
\label{quadraticLDiff}
\eea
These still allow for arbitrary values of $g_1$ and $g_2$, and
\bea
&&g_3=-\left(2g_1+g_2\right)\nonumber\\
&&g_4=\dfrac{2g_1+g_2}{ 2}\nonumber\\
&&g_5=-\dfrac{g_2}{ 2}
\eea
In the second place we consider invariance under  {\em LTDiff}, which imposes
\be
g_3+ 2 g_4=0
\ee
\par
Finally for {\em LWeyl} symmetry, the variations read (multiplying by $\kappa/2$ again for simplicity)
\bea
&&\d{\cal O}_1=2 \omega \, \Box \pd^\a\pd^\b h_{\a\b}\nonumber\\
&&\d{\cal O}_2=\omega \, \Box^2 h+ 4 \omega \, \Box \pd^\a\pd^\b h_{\a\b}\nonumber\\
&&\d{\cal O}_3= 2 \omega \, \Box \pd^\a\pd^\b h_{\a\b}\nonumber\\
&&\d{\cal O}_4=2 \omega \, \Box^2 h\nonumber\\
&&\d{\cal O}_5=8 \omega \, \Box^2 h
\label{Weylvar}
\eea
so that the action is invariant under such tranformations whenever
\bea
&2g_1+ 4 g_2+ 2 g_3=0\nonumber\\
&g_2+2g_4+ 8 g_5=0
\label{WeylConditions}
\eea

Now it is interesting to combine {\em LDiff} and {\em LWeyl}. In the case of dimension 4 operators, actions which are invariant under both symmetries do not exist. For dimension 6 operators, we can have {\em LWDiff}  invariant theories as long as the coupling constants are constrained to have the following relations
\be
g_1 =  g_2, \quad g_3 = -3g_2 , \quad g_4 = \dfrac{3}{2} g_2, \quad g_5 = -\dfrac{g_2}{2}
\ee
These actions with {\em LWDiff} invariance are obtained as the weak field limit of the following quadratic theories
\be
L=\sqrt{g} \, \left[ \a\, R_{\a\b\g\d}^2+\left[-4 \a+6  g_2  \right] R_{\m\n}^2+ \left(\a - 2 g_2 \right) R^2\right]
\ee
Note that the term $\sqrt{g}$ is immaterial at the order we are working. The weak field expansion of the quadratic invariants is worked out in the appendix \ref{A}.  For $n=4$ spacetime dimension these theories can be rewritten as
\be
L=\sqrt{g} \, \left(\a - 3 g_2 \right)\,E_4+\sqrt{g} \, 3 g_2 \, W_4
\ee 
where $E_4$ is the four-dimensional Euler density
\be
E_4\equiv R_{\a\b\g\d}^2-4\, R_{\a\b}^2+ R^2
\ee
It can be easily checked that the weak field expansion of $E_4$ around Minkowski spacetime vanishes. This is the origin of the arbitrary coefficient $\a$ in the above expression.
The quantity $W_4$ is the square of the 4-dimensional Weyl tensor
\be
W_4 \equiv R_{\a\b\g\d}^2-2\, R_{\a\b}^2+{1\over 3}\, R^2
\ee
The actions which are precisely proportional to the Weyl squared tensor are the ones with $ 3 g_2 = \a  $. Nevertheless,
at the linear level, the Euler density does not contribute, so that all the solutions will effectively correspond to actions proportional to $W_4$.
\par

Also quite interesting are those actions that are {\em LWTDiff} invariant; that is {\em LWeyl} invariant, but {\em LDiff} invariant under transverse diffeomorphisms only. They are characterized by
\bea
&&g_1=-g_3-2 g_2= 6 g_4+16 g_5\nonumber\\
&&g_2=-2 g_4- 8 g_5\nonumber\\
&&g_3=-2 g_4
\label{consWTDiff}
\eea
The most general quadratic  {\em WTDiff} invariant lagrangian is the one obtained by Weyl transforming the metric in the quadratic action with $\tilde{g}_{\mu\nu} = g^{-1/4} g_{\m\n}$ (this transformation ensures {\em TDiff} and automatically introduces a Weyl invariance). The expansion around flat spacetime reads
\bea
& \sqrt{\tilde{g}} \, (\a \tilde{R}_{\m\n\r\s} \tilde{R}^{\m\n\r\s}+ \b  \tilde{R}_{\m\n} \tilde{R}^{\m\n} + \g \tilde{R}^2)  = \left(\a+{\b\over 2}+\g\right)\,{\cal O}_1 +\dfrac{1}{2}(\a-\g)\,{\cal O}_2 \nonumber\\
&-\left(2 \a +{\b\over 2}\right)\,{\cal O}_3+ \left(\a+{\b\over 4}\right)\,{\cal O}_4-\left(\dfrac{20\a+4\b -4\g}{64}\right)\,{\cal O}_5 +O(h^3) \nonumber \\
\eea
The weak field limit of these theories automatically satisfies the constraints needed for {\em LWTDiff} \eqref{consWTDiff}. The precise form of these theories after the Weyl transformation \eqref{Weylquadratic} is shown in the appendix. 
\subsubsection{Scale and conformal invariance}
{\em Scale invariance} is now lost with the assingment given to $h_{\m\n}$ (conformal weight one). If we have a theory incorporating dimension 6 operators only, it is possible to recover scale invariance, just by making the graviton inert (conformal weight 0). It is plain that this does not hold when we have both dimension 4 and dimension 6 operators in the theory.

On the other hand, the {\em conformal invariance} demands as usual, tracelessness of the (metric, or Rosenfeld) energy-momentum tensor. 
In this case, the   energy-momentum tensor takes the form
\bea
T_{\m\n}&=&2 g_1\{h_{\m\l} \pd_\n \pd^\l\pd^\a \pd^\b h_{\a\b}+h_{\n\l} \pd_\m \pd^\l\pd^\a \pd^\b h_{\a\b}\}\nonumber\\
&&+g_2\{h_{\l \m} \pd_\n \pd^\l \Box h+h_{\l \n} \pd_\m \pd^\l \Box h + h_{\a\b}\pd^{\a}\pd^\b\pd_{\m} \pd_{\n} h+h_{\a\b}\pd^{\a}\pd^{\b}\Box h_{\m\n}     \}\nonumber\\
&&+g_3\{ h_{\a \m} \pd_\n \pd_\b \Box h^{\a\b} +h_{\a \n} \pd_\m \pd_\b \Box h^{\a\b} + h_{\a\m} \pd^\a \pd^\b \Box h_{\b \n } + h_{\a \b} \pd^\a \pd_\l \pd_{\m} \pd_{\n} h^{\l \b }\}\nonumber\\
&&+2g_4\{ h_{\a\b} \pd_{\m}\pd_{\n} \Box h^{\a\b} + h_{\l\m} \Box^2 h^\l_\n         \}  + 2g_5 \{  h \pd_{\m} \pd_{\n} \Box h +  h_{\m\n}  \Box^2 h \}-{1\over 2}\,L\eta_{\m\n}
\eea
and the trace reads
\be
T=2\,L
\ee
The eom read
\bea
\dfrac{\d S}{ \d h^{\a\b}}&=& 2g_1\pd_\a \pd_\b \pd^\m \pd^\n h_{\m\n}+ g_2\pd_\a\pd_\b \Box h+ g_2\eta_{\a\b}\pd^\m\pd^\n \Box h_{\m\n}
+g_3\pd_\b \pd^\l \Box h_{\a\l}+g_3\pd_\a \pd^\l \Box h_{\b\l}+\nonumber\\
&&+2g_4\Box^2 h_{\a\b}+2g_5\eta^{\a\b}\Box^2h\nonumber\\
\eea
and they imply that, on-shell, the lagrangian indeed vanishes, $L=0$ (up to total derivatives). 

In order to study the virial in detail, let us start from the specific form of the lagrangian
\bea
L&=& 2 \left\lbrace  g_1 \left(\pd^\a  \pd^\g h_{\a\b} \pd^\b  \pd^\d h_{\g\d}+\pd^\a   h_{\a\b} \pd^\b  \pd^\g\pd^\d h_{\g\d}+\pd^\g   h_{\a\b} \pd^\b  \pd^\a\pd^\d h_{\g\d}\right)+ \right. \nonumber\\
&& \left. +\frac{g_2}{2} \left(\pd^\a\pd^\b  h_{\a\b}\Box h+\pd^\a\pd^\b h \Box h_{\a\b}+2\pd^\b  h_{\a\b}\pd^\a\Box h+2\pd^\b h \Box\pd^\a h_{\a\b}\right)+ \right. \nonumber\\
&&\left. +g_3\left( \pd^\a  \pd^\b h_{\a\l} \Box  h_{\b}^\l+ \pd^\a h_{\a\l}  \pd^\b \Box  h_{\b}^\l +\pd^\b h_{\a\l}  \pd^\a \Box  h_{\b}^\l\right)+ \right. \nonumber\\
&&\left. +g_4 \left(\Box h_{\a\b}\Box h^{\a\b}+2\pd^\m h_{\a\b}\pd_\m\Box h^{\a\b}\right)+ \right. \nonumber\\&& \left. +g_5\left(\Box h\Box h+2\pd^\m h\pd_\m\Box h\right) \right\rbrace
\eea
Now it is a simple matter to show that on-shell, $L=\pd_\m\pd_\n\s^{\m\n}$ with
\bea
\s^{\m\n}&=& 2 \left\lbrace \frac{g_1}{2}\left(h^{\m\a} \pd_\a \pd^\b h_{\b}^\n+h^{\n\a} \pd_\a \pd^\b h_{\b}^\m\right)+\frac{g_2}{2}\left(h^{\m\n}\Box h+h\Box h^{\m\n}\right)+ \right. \nonumber\\
&& \left. +\frac{g_3}{2}\left(h^{\m\l}\Box h^\n_\l+h^{\n\l}\Box h^\m_\l\right)+g_4\eta^{\m\n} h_{\a\b}\Box h^{\a\b}+g_5\eta^{\m\n} h\Box h \right\rbrace
\eea
This result is somewhat puzzling, because we have already indicated that this theory is {\em not}  even scale invariant with the standard assignment of conformal weight for the graviton field (namely 1). The result is however logical if we remember that the low energy of the Weyl squared lagrangian is of this form. The theory containing both dimension 4 as well as dimension 6 operators, should not be conformal however.
 This fact can be easily understood from a simpler example, namely a scalar lagrangian, where all complications of indices can be avoided. Consider then the lagrangian
\be
L' = \a \phi \Box \phi +\dfrac{\b}{M^2} \phi \Box^2 \phi
\label{dim46}
\ee
which is equivalent to (up to total derivatives)
\be
L = -\a \, \pd_\m \phi \pd^\m \phi - \dfrac{\b}{M^2} \left(4 \pd_\l \phi \Box \pd^\l \phi + 2 \pd^\m \pd_\l \phi \pd_\m \pd^\l \phi +  \Box \phi \Box \phi \right) = -\a L_1 - \dfrac{\b}{M^2}L_2
\ee
This is our starting point. The eom read
\be
{\d S\over \d\phi}= \a \Box \phi + \dfrac{\b}{M^2} \Box^2 \phi = 0 
\ee
and the energy-momentum tensor
\bea
T_{\m\n} &=& -\a \Big( \pd_\m \phi \pd_\n \phi - \dfrac{1}{2} \pd_\l \phi \pd^\l \phi \,  \eta_{\m\n} \Big) - \dfrac{\b}{M^2}  \Big(4 \pd_\l\phi \pd_\m\pd_\n \pd^\l \phi + 4 \pd_\m \phi \Box \pd_\n \phi + 4 \pd_\m \pd_\l \phi \pd_\n \pd^\l \phi \nonumber \\
&& + 2 \pd_\m\pd_\n \phi \Box \phi - \dfrac 1 2\left(4 \pd_\l \phi \Box \pd^\l \phi + 2 \pd^\s \pd_\l \phi \pd_\s \pd^\l \phi +  \Box \phi \Box \phi \right) \eta_{\m\n} \Big)
\eea
The trace of the above reads
\be
T = -\a  \left(1 - \dfrac{n}{2}\right) L_1 -  \dfrac{\b}{M^2}\left(2 - \dfrac{n}{2}\right) L_2 = \a   L_1
\ee
Even if we are working in $n=4$, we leave $n$ arbitrary to maintain the second piece and illustrate the point we want to make. 
Note that this is not proportional to the total lagrangian, because the trace counts the number of derivatives. 
We can rewrite the trace as 
\be
T = - \dfrac{\a}{2}  \left(1 - \dfrac{n}{2}\right) \left(\Box \phi^2 - 2 \phi \Box \phi \right) -  \dfrac{\b}{2 M^2}\left(2 - \dfrac{n}{2}\right) \left(\Box^2 \phi^2 - 2 \phi \Box^2 \phi \right)
\ee
which fails to be a total derivative when both $\a$ and $\b$ are nonvanishing, because
\be
1-{n\over 2}\neq 2-{n\over 2}
\ee
\par
Note that this is true even if there are {\em WTDiff} (that is {\em TDiff} and {\em Weyl} invariant) theories linear as well as quadratic in the Riemann tensor. The weak field limit of those Weyl invariant theories fails to be conformal invariant.
\subsection{Dimension 5 and dimension 6 operators (with 2 derivatives)}
Next, we study dimension 5 and dimension 6 operators containing just two derivatives, so that they come from the weak field limit of theories linear in the curvature, when expanded to higher orders in the perturbation. In the previous sections, operators coming from the lowest (non-trivial) order of gravitational actions were analyzed. In that cases, the lowest order of {\em (T)Diff} and Weyl variations was enough to obtain the conditions for those actions to be invariant under such symmetries. In this case, however, different orders of the expansion are needed because of the two orders involved in the field variations
\bea
&&\delta_D  (\kappa \,h_{\m \nu}) = \pd_\m \xi_\n +\pd_\n \xi_\m + \kappa \, \mathcal{L}_{\xi} h_{\m \nu} \nonumber \\
&&\delta_W( \kappa \, h_{\m \nu}) = 2 \omega \eta_{\m\n} + 2 \kappa \, \omega h_{\m \nu}
\label{variations}
\eea
This translates into dimension 4 operators mixing with dimension 5 ones, and dimension 5 operators with dimension 6 ones. 
\par
A full list of the independent dimension 5 and dimension 6 operators (containing 2 derivatives) can be found in appendix \ref{56}. The most general dimension 5 lagrangian with such operators reads
\be
\mathcal{L}_{5 \pd\pd} = \kappa  \sum_i^{14} a_i \mathcal{N}_i
\ee
Again, the only diffeomorphism invariant combination corresponds to $\sqrt{g} R$, in this case, to the order $O(\kappa^3)$ expansion of it
\bea \left(\dfrac{1}{\kappa^2}\sqrt{g}R \right)_{O(\kappa^3)}&&= \kappa \left\lbrace-\frac{1}{4}\mathcal{N}_{1}-\frac{1}{2}\mathcal{N}_{2}+\frac{1}{4}\mathcal{N}_{3}-\frac{1}{2}\mathcal{N}_{4}+\frac{1}{2}\mathcal{N}_{5}-\frac{1}{4}\mathcal{N}_{6}- \right.\nonumber\\
&&\left.-\frac{1}{4}\mathcal{N}_{8}-\frac{1}{8}\mathcal{N}_{9}-\frac{1}{4}\mathcal{N}_{11}+\frac{1}{8}\mathcal{N}_{12}+\frac{3}{16}\mathcal{N}_{13}-\frac{1}{16}\mathcal{N}_{14} \right\rbrace\eea
This piece then combines with the previous order of the expansion to attain diffeomorphism invariance
\be
\left(\sqrt{g}R \right)_{O(\kappa^2)}\Big|_{\d ( \kappa \, h_{\m \nu}) = \pd_\m \xi_\n +\pd_\n \xi_\m}+\left(\sqrt{g}R \right)_{O(\kappa^3)}\Big|_{\d ( \kappa \, h_{\m \nu}) =\kappa \, \mathcal{L}_{\xi} h_{\m \nu} } = 0
\ee
We can also look for the most general Lorentz and Weyl invariant lagrangian built with this kind of operators. Again, we need the dimension 4 operator part that will contribute with the $O(h)$ piece of the Weyl variation, which already has two arbitrary constants appearing in it \eqref{4dWeyl}. Taking that piece into account, the most general Weyl invariant lagrangian up to dimension 5 operators reads
\bea
 \mathcal{L}_{W_{5D}}&&=\frac{c_1}{4} \partial_\l h_{\m\n}\partial^\l h^{\m\n}-\frac{c_2}{2} \partial_\m h^{\m\l}\partial_\n h^{\n}_{~\l}+\frac{c_2}{4}\partial_\n h^{\m\n}\partial_\m h-\frac{2c_1+c_2}{32}\partial_\l h\partial^\l h +\nonumber \\
 && + \kappa \left\lbrace a_{1}\mathcal{N}_{1}+a_{2}\mathcal{N}_{2}+a_{3}\mathcal{N}_{3}+4a_{1}\mathcal{N}_{4}+a_{5}\mathcal{N}_{5}+\left(a_{1}-2a_{2}-4a_{3}+\frac{c_2}{4}\right)\mathcal{N}_{6}+ \right. \nonumber\\
 &&+\left(4a_{1}-4a_{2}-a_{5}\right)\mathcal{N}_{7}+\left(-a_{1}+2a_{2}-\frac{c_2}{4}\right)\mathcal{N}_{8}+a_{9}\mathcal{N}_{9}+\frac{1}{16}\left(2a_{1}-4a_{2}-4a_{9}+c_2\right)\mathcal{N}_{10}+\nonumber\\
 &&+a_{11}\mathcal{N}_{11}+\frac{1}{16}\left(-4a_{1}-8a_{11}+8a_{2}+16a_{3}+2c_1-c_2\right)\mathcal{N}_{12}-\frac{1}{4}\left(a_{11}+a_{9}\right)\mathcal{N}_{13}+\nonumber\\
 && \left. +\frac{1}{32}\left(a_{1}+4a_{11}-2a_{2}-8a_{3}+2a_{9}-c_1\right)\mathcal{N}_{14} \right\rbrace
 \label{W5D}
\eea
Let us insist on the fact that both pieces are needed so that Weyl invariance is attained, that is, up to certain order in the expansion, the previous order is needed for the computation of the invariance conditions. Thanks to the mixing of the different orders, more freedom is avalaible to attain invariance under the studied symmetries. 
\par
It is straightforward to see that  $WTDiff$ \eqref{wtdiff} is a particular case of this general Weyl invariant lagrangian, with the constants fixed to
\be
a_{1}=-\frac{1}{8}, \, a_{2}=-\frac{1}{4}, \, a_{3}=\frac{3}{32}, \,  a_{5}=\frac{1}{2},  \, a_{9}=0,  \, a_{11}=-\frac{1}{4},  \, c_1= c_2=-1
\ee
\par
Accordingly, the most general dimension 6 lagrangian with two derivatives can be written as
\be
\mathcal{L}_{6 \pd\pd} = \kappa^2 \sum_i^{38} b_i \mathcal{K}_i
\ee
In this case, the expansion of $\left(\sqrt{g}R \right)_{O(\kappa^4)}$ reads
\bea \left(\dfrac{1}{\kappa^2}\sqrt{g}R \right)_{O(\kappa^4)}&=& \kappa^2 \left\lbrace  \frac{1}{4}\mathcal{K}_{1}-\frac{1}{8}\mathcal{K}_{3}+\frac{1}{2}\mathcal{K}_{4}-\frac{1}{4}\mathcal{K}_{5}-\frac{3}{16}\mathcal{K}_{6}+\frac{1}{16}\mathcal{K}_{7}+\frac{1}{32}\mathcal{K}_{8}-\frac{1}{2}\mathcal{K}_{9}+\frac{1}{2}\mathcal{K}_{11}+ \right. \nonumber\\
&&+\frac{1}{8}\mathcal{K}_{12}-\frac{1}{8}\mathcal{K}_{13}-\frac{1}{2}\mathcal{K}_{14}+\frac{1}{4}\mathcal{K}_{16}-\frac{1}{8}\mathcal{K}_{17}+\frac{1}{4}\mathcal{K}_{20}-\frac{1}{16}\mathcal{K}_{21}+\frac{1}{12}\mathcal{K}_{22}-\nonumber\\
&&-\frac{1}{16}\mathcal{K}_{23}+\frac{1}{4}\mathcal{K}_{25}-\frac{1}{8}\mathcal{K}_{26}-\frac{1}{8}\mathcal{K}_{27}+\frac{1}{32}\mathcal{K}_{28}-\frac{1}{8}\mathcal{K}_{29}+\frac{3}{32}\mathcal{K}_{30}-\frac{1}{96}\mathcal{K}_{31}+\nonumber\\
&& \left. +\frac{1}{2}\mathcal{K}_{32}-\frac{1}{2}\mathcal{K}_{33}-\frac{1}{4}\mathcal{K}_{35}-\frac{1}{16}\mathcal{K}_{38}\right\rbrace\eea

Finally, let us analyze the most general Weyl invariant lagrangian up to dimension 6 operators. We have different pieces apearing in it. First of all, it contains the pieces up to dimension 5 that were computed in this section \eqref{W5D}, together with the dimension 6 piece of two derivative operators, that combine with specific coefficients so that Weyl invariance  is attained. Moreover, we have another Weyl invariant combination coming from dimension 6 operators containing four derivatives \eqref{WeylConditions}. Taking everything into account, the most general Weyl invariant lagrangian up to dimension six operators is shown in appendix \ref{56}.
\subsection{Interaction terms}
It would appear quite intuitive to think that there are no potential terms invariant under either {\em Diff} or Weyl invariance. This is based in our GR intuition, but let us get rid of those prejudices and carry on with our perturbative analysis.
It is easy to systematize the perturbative expansion.  Up to quartic interactions we have the monomials
\begin{equation*}
\begin{aligned}[c]
&{\cal M}_1\equiv \, h_{\a\b} h^{\a\b}\nonumber\\
&{\cal M}_2\equiv \, h^2\nonumber\\
& \quad 
\end{aligned} \quad \quad 
\begin{aligned}[c]
& {\cal J}_1\equiv h^{\a\b} h_{\b\g}  h^{\g}_\a \nonumber\\
& {\cal J}_2\equiv h^{\a\b} h_{\a\b} h \nonumber\\
& {\cal J}_3 \equiv h^3\nonumber\\
\end{aligned} \quad\quad 
\begin{aligned}[c]
&{\cal Q}_1\equiv h^{\a\b} h_{\b\g}  h^{\g\d} h_{\d\a}\nonumber\\
&{\cal Q}_2\equiv \left(h_{\m\n} h^{\m\n}\right)^2\nonumber\\
& \quad 	
\end{aligned} \quad \quad 
\begin{aligned}[c]
&{\cal Q}_3\equiv h^4\nonumber\\
&{\cal Q}_4\equiv h^2 h_{\a\b}h^{\a\b}\nonumber\\
&{\cal Q}_5\equiv h h_{\a\b}h^{\b\g}h_\g^\a \nonumber \\
\end{aligned}
\end{equation*}
so that the most general potential  up to dimension four will read
\be
V(h_{\m\n}) =    m M^2  h + \sum_{i=1}^2 m_{i}^2{\cal M}_i +\sum_{a=1}^{a=3} b_a {\cal J}_a +\sum_{a=1}^{a=5} \l_a {\cal Q}_a+\ldots
\ee
\par
We want to analyze the invariance under diffeomorphisms as if this $h_{\m\n}$ corresponds to the perturbation of the metric around flat spacetime \eqref{variations}, but we take another energy scale $M$ instead of $\kappa$. The crucial point is that owing to the fact that the {\em Diff} variations contain an order zero piece and an order one piece in the perturbation, each order in the perturbative expansion of the variation of the potential contributes to both the lower and upper orders. Up to total derivatives and dimension four operators, it can be seen that the following interaction lagrangian is diffeomorphism invariant
\bea
&V^D(h_{\m\n})= m M^2 \, h + \dfrac{m M}{4 } \, \left({\cal M}_2-2 {\cal M}_1 \right) + m  \, \left(\dfrac{1}{3}  {\cal J}_1 -\dfrac{1}{4}  {\cal J}_2+\dfrac{1}{24}  {\cal J}_3 \right) +  \nonumber\\
& +\, \dfrac{m}{M} \, \left( -\dfrac{1}{4} {\cal Q}_1+\dfrac{1}{16}{\cal Q}_2 +\frac{1}{192}{\cal Q}_3 
-\dfrac{1}{16} {\cal Q}_4+\dfrac{1}{6} {\cal Q}_5\right)+\ldots
\eea
In fact this is an iterative process, each term in the expansion determining the following. The final potential contains infinite terms depending on just one arbitrary constant with dimensions of mass, $m$. In fact this is exactly the weak field expansion of $m M^4 \left(\sqrt{g}-1\right)$.
\par
At this point our GR intuition strikes back and asks whether this is not precisely the expansion of  the cosmological constant term. (In fact they do not quite fit).
\par
Concentrating in the quadratic terms
\be
V_2^D\equiv m M^2 h+{m M \over 4}    h^2-{m M \over 2} h_{\a\b}^2=-{M m \over 2} \left(h_{\a\b}+M \eta_{\a\b}\right)^2+{m M \over 4}   \left( h+4 M\right)^2-2 m M^3
\ee
Not knowing anything on GR we would say that there is spontaneous symmetrÁy breaking in the system and the ground state has shifted from
$h_{\m\n}=0$ to $h_{\m\n}= -M \eta_{\m\n}$, leaving behind a vacuum energy
\be
V_0^D\equiv -2 m M^3
\ee
Fluctuations around the new vacuum state
\be
h_{\m\n}\equiv  -M \eta_{\m\n}+H_{\m\n}
\ee
are damped (provided $m M <0$) with a quadratic term
\be
V_2^D+2 m M^3=-{m M\over 2}  H_{\a\b}^2+{m M \over 4} H^2
\ee
as is not positive semidefinite except for traceless $H_{\a\b}^T$ when $m M<0$. In order to reach a definite conclusion on positivity, higher order terms should be considered. To the extent that this is related to the weak field expansion of $m M^2 \sqrt{g}$,  we expect it to have a definite sign however. 
\par
Similar reasoning  as in the previous paragraph leads to a Lorentz and Weyl invariant potential
\bea
V^W(h_{\m\n})&=& m M^2 \, h + m_1^2 \, {\cal M}_1 - \dfrac 1 8  \left( m M +2 m_1^2  \right)\, {\cal M}_2+ b_1 {\cal J}_1  + \dfrac{1}{4}  \left(-2 { m_1^2\over M} -3 b_1 \right) {\cal J}_2 + \nonumber \\
&&+ \dfrac{1}{48}  \left(m + 6 {m_1^2\over M} + 6 b_1 \right)   {\cal J}_3 + \kappa \, m \, \lambda_1 {\cal Q}_1+ \lambda_2{\cal Q}_2  + \dfrac{1}{256} \left( -{ m\over M} -12 { m_1^2\over M^2} - 24 { b_1\over M} -  \right. \nonumber \\ \nonumber \\
&&\left. - 12 \l_1 + 16 \l_2 \right){\cal Q}_3+ \dfrac{1}{16}\left( 3 { m_1^2\over M^2} + 9 {b_1\over M} + 6 \l_1 -8\l_2\right)  {\cal Q}_4 - \left(\dfrac{3}{4}  {b_1\over M} + \l_1  \right) {\cal Q}_5
\eea
In this case we have more freedom as more arbitrary constants appear with each order of the perturbative expansion.
The quadratic piece can be written as
\be
V_2^W=m_1^2\left(h_{\a\b}-M \eta_{\a\b}\right)^2-{m M+ 2 m_1^2\over 8} \left(h- 4 M\right)^2+ 2 m M^3
\ee
Fluctuations around the minimum of the potential 
\be
h_{\m\n}= M \eta_{\m\n}+ H_{\m\n}
\ee
behave as
\be
V_2^W =m_1^2 H_{\m\n}^2-{m M+ 2 m_1^2\over 8} H^2+ 2 m M^3
\ee
which again is positive semisefinite only for traceless $H_{\a\b}^T$ or else for pure trace when $m M <0$ as
\be
V_2^W =m_1^2 {H_{\m\n}^T}^2- \dfrac{mM}{8}H^2+ 2 m M^3
\ee
\subsection{Global Weyl invariance}
There is another symmetry that can be studied in this context, which is global Weyl invariance, that is, when the Weyl scaling factor is just a constant
\be
\delta g_{\m\n} = \Omega^2 g_{\m\n} \, ,  \quad \partial_\m \Omega = 0
\ee
When we expand the metric around flat spacetime, $g_{\m\n} = \eta_{\m\n} + \kappa h_{\mu\nu}$, the linearized variation of the quantum fluctuation reads
\be
\d h_{\m\n} = 2 \omega \left(\dfrac{1}{\kappa} \eta_{\m\n} + h_{\m\n}\right)
\ee
In the case of global (rigid) Weyl invariance where $\omega$ is constant, the variations of the operators quadratic in the fields have to be computed taking into account both terms in the above (that is, the linear order in the quantum field). If we just took the first piece, proportional to the Minkowski metric, all the variations computed in \eqref{Weylvar1} and \eqref{Weylvar} would just be total derivatives, which have been neglected in this work. \\

In order to illustrate this point, let us take two simple actions. We know that the Einstein Hilbert action is not globally Weyl invariant in four dimensions,   
\be
-\dfrac{1}{2 \kappa^2} \int d^4 x \, \d(\sqrt{g} R) = -\dfrac{1}{2 \kappa^2} \left( 2 \omega  \, \int d^4x \, \sqrt{g} R  \right)
\ee
On the other hand, we can take the simplest quadratic action which is invariant in four dimensions
\be
\int d^4 x \, \d(\sqrt{g} R^2) = 0
\ee

These equalities have to be true order by order in the perturbation of the metric. In this case, the quadratic order in the variation together with the linear order in the Weyl variation, combines with the third order of the perturbation in the action and the lowest order in the Weyl variation. These terms are going to be of order $O(\kappa^2)$ and have to match exactly the $O(\kappa^2)$ part of the rhs of the equation. Namely,

\be
  \int d^4 x \,\left[(\sqrt{g} R)^{O(\kappa^2)}\Big|_{\d h_{\mu\nu}= 2 \omega h_{\mu\nu} } +(\sqrt{g} R)^{O(\kappa^3)}\Big|_{\d h_{\mu\nu}=  2 \omega \frac{\eta_{\mu\nu} }{\kappa} }\right]=  \left( 2 \omega  \,  \int d^4x \, (\sqrt{g} R )^{O(\kappa^2)} \right)
\label{LinearGlobalWeyl}
\ee
The quadratic expansion of the Einstein Hilbert action reads
\be
-\dfrac{1}{2\kappa^2}\int d^4x \,  (\sqrt{g} R )^{O(\kappa^2)} =-\dfrac{1}{2} \int d^4x \,\left\lbrace -\dfrac{1}{2} \partial_\m h^{\m\n}\partial_\n h + \dfrac{1}{2} \partial_\m h^{\m\n}\partial_\r h^\r_\n   + \dfrac{1}{4} \partial_\m h \partial^\m h- \dfrac{1}{4} \partial_\r h^{\m\n}\partial^\r h_{\m\n} \right\rbrace
\ee
It is straightforward to see that taking the variation $\d h_{\mu\nu}= 2 \omega h_{\mu\nu}$ we get
\be
\int d^n x \, (\sqrt{g} R)^{O(\kappa^2)}\Big|_{\d h_{\mu\nu}= 2 \omega h_{\mu\nu} } = 4 \omega \int d^n x \,  (\sqrt{g} R )^{O(\kappa^2)} 
\ee
In order to compute the other piece contributing to $O(\kappa^2)$ we need the third order of the expansion of the Einstein Hilbert action which contains terms with three quantum fields $h_{\m\n}$ and two derivatives, which are shown in the appendix \ref{56}. Once we have this expansion, we perform the Weyl variation $\d h_{\mu\nu}= 2 \omega \frac{\eta_{\mu\nu}}{\kappa}$ yielding
\be
\int d^n x \, (\sqrt{g} R)^{O(\kappa^2)}\Big|_{\d h_{\mu\nu}= 2 \omega \frac{\eta_{\mu\nu} }{\kappa} } =  -2 \omega \int d^n x \,  (\sqrt{g} R )^{O(\kappa^2)} 
\ee
Adding the two contributions
\be
\int d^n x \,\left[(\sqrt{g} R)^{O(\kappa^2)}\Big|_{\d h_{\mu\nu}= 2 \omega h_{\mu\nu} } +(\sqrt{g} R)^{O(\kappa^3)}\Big|_{\d h_{\mu\nu}=  \frac{\eta_{\mu\nu} }{\kappa} }\right]= 2 \omega  \, \int d^n x \, (\sqrt{g} R )^{O(\kappa^2)}
\ee
which precisely yields the right hand side of \eqref{LinearGlobalWeyl} for $n=4$. We can see that in $n=2$ the Einstein Hilbert action is globally Weyl invariant (as well as locally). In fact this is basically the reason all two-dimensional metrics are conformally flat.\\

In the case of the quadratic action we have
\be
\int d^4 x \,\left[(\sqrt{g} R^2)^{O(\kappa^2)}\Big|_{\d h_{\mu\nu}= 2 \omega h_{\mu\nu} } +(\sqrt{g} R^2)^{O(\kappa^3)}\Big|_{\d h_{\mu\nu}= 2 \omega \frac{\eta_{\mu\nu} }{\kappa} }\right]= 0
\label{QuadraticGlobalWeyl}
\ee
where
\be
\int d^4 x \,(\sqrt{g} R^2)^{O(\kappa^2)} =\kappa^2  \int d^4 x \, \left\lbrace \partial_\m \partial_\n h^{\m\n} \partial_\r \partial_\s h^{\r \s} -2 \partial_\m \partial_\n h^{\m\n} \Box h + \Box h \Box h   \right\rbrace
\ee
As before, taking the Weyl variation proportional to the quantum field is straightforward and it yields
\be
\int d^n x (\sqrt{g} R^2)^{O(\kappa^2)}\Big|_{\d h_{\mu\nu}= 2 \omega h_{\mu\nu} } = 4 \omega \int d^n x  (\sqrt{g} R^2)^{O(\kappa^2)}
\ee
For the other piece, we need the third order variation of the quadratic action which can be easily computed. After performing the Weyl transformation on the quantum field, $\d h_{\mu\nu}= 2 \omega \frac{\eta_{\mu\nu} }{\kappa}$, we get
\be
\int d^n x \,(\sqrt{g} R^2)^{O(\kappa^3)}\Big|_{\d h_{\mu\nu}= 2 \omega \frac{\eta_{\mu\nu} }{\kappa} } = -4 \omega \int d^n x  (\sqrt{g} R^2)^{O(\kappa^2)}
\ee
Summing both contributions, 
\be
\int d^4 x \,\left[(\sqrt{g} R^2)^{O(\kappa^2)}\Big|_{\d h_{\mu\nu}= 2 \omega h_{\mu\nu} } +(\sqrt{g} R^2)^{O(\kappa^3)}\Big|_{\d h_{\mu\nu}= 2 \omega \frac{\eta_{\mu\nu} }{\kappa} }\right]= 0
\ee
\section{Non-local extensions}
There is a permanent temptation to avoid the K\"allen-Lehman spectral theorem (which states that the price to pay for having propagators that fall off at euclidean infinity faster than  $k^{-2}$ is to have negative norm states) by considering non-local theories. For example in \cite{Biswas, Modesto} a non-local generalization of the dimension 4 operators has been considered, namely
\bea
&&\frak{O}_1\equiv -{1\over 4}\,h_{\a\b}\left[c_1\left({\Box\over M^2}\right)\,\Box\right]\,h^{\a\b}\nonumber\\
&&{\frak O}_2\equiv {1\over 2} h_{\a\b}\left[c_2\left({\Box\over M^2}\right)\,\pd^\a\pd^\g\right]\,h_{\g}^{\b}\nonumber\\
&&{\frak O}_3\equiv-{1\over 2}\, h\left[ c_3\left({\Box\over M^2}\right)\,\pd^\g\pd^\d \right]h_{\g\d}\nonumber\\
&&{\frak O}_4\equiv {1\over 4}\,h\left[ c_4\left({\Box\over M^2}\right)\, \Box\right] h\nonumber\\
&&{\frak O}_5\equiv h_{\a\b}\left[c_5\left({\Box\over M^2}\right)\,{\pd^\a\pd^\b\pd^\g\pd^\d\over \Box}\right]\,h_{\g\d}
\eea
so that the general lagrangian of this type will be
\be
L=\sum_{i=1}^{i=5}{\frak O}_i
\ee
(where $c_i(z)$ are analytic functions with dimensionless argument).
The five functions $c_i(z),\,i=1\ldots 5$ (which are assumed to include the corresponding coupling constants) characterize the theory. The constants put in front are such that the {\em LDiff} Fierz-Pauli theory corresponds to
\bea
&{\cal O}_i\leftrightarrow {\frak O}_i\quad (i=1\ldots 5)\nonumber\\
&g_1=g_2=g_3=g_4=1\nonumber\\
&g_5=0
\eea
The correspondence with the dimension 6 operators in \eqref{seis} is as follows
\bea
&&c_i(z)=z\quad (i=1\ldots 5)\nonumber\\
&&{\cal O}_1\leftrightarrow M^2\,{\frak O}_5\nonumber\\
&&{\cal O}_2\leftrightarrow -2 M^2 \,{\frak O}_3\nonumber\\
&&{\cal O}_3\leftrightarrow 2 M^2 \,{\frak O}_2\nonumber\\
&&{\cal O}_4\leftrightarrow -4 M^2 \,{\frak O}_1\nonumber\\
&&{\cal O}_5\leftrightarrow 4 M^2\,{\frak O}_4
\eea
i.e. $g_1=M^2 c_5(z)$, $g_2=-2 M^2 c_3(z)$, $g_3=2 M^2 c_2(z)$,$g_4=-4 M^2 c_1(z)$ and $g_5=4 M^2 c_4(z)$, in such a way that the conditions for {\em LDiff} invariance now translate into 
\bea
&&c_2(z) - c_3(z) + c_5(z) =0 \nonumber \\
&&4 c_4(z) -c_3(z) = 0 \nonumber \\
&&c_2(z) -4c_1(z) = 0\label{LD}
\eea
It is claimed in \cite{Biswas} that the theory is {\em ghost-free} provided that
\bea
&&c_1(z)=c_2(z)\nonumber\\
&&c_3(z)=c_4(z)\nonumber\\
&&c_5(z)=2\left(c_3(z)-c_2(z)\right)\label{Bis}
\eea
and the function $c_1(z)$ is chosen as an entire function, such as
\be
c_1\,(z)\equiv e^{-z}
\ee
Note that both constraints, \eqref{LD} and \eqref{Bis} are different and incompatible.

It is well-known, however, that non-local theories suffer from unitarity and causality problems, some of those can be sometimes hidden uunder the rug of experimental precision of the measurements \cite{Tomboulis}. However, in order to do that, the theory needs to be {\em quasi-local}, which means that the corresponding function has got to have bounded support, which seems to contradict other conditions. It is not clear at all that a consistent solution exists.
\par
Outstanding problems in this respect according to \cite {Asorey} are first and foremost, the fact that the presence of the exponential damping factor in the propagator prevents analytic continuation from the riemannian theory to the lorentzian one, owing to the essential singularities in the complex energy plane. It must be stressed, however that such an analytic continuation is problematic in {\em any} theory involving the gravitational field.  Another argument is that  none of the theories proposed so far complies with reflexion-positivity, which is believed to be an essential requirement in order to get a consistent quantum field theory.

\section{Conclusions}
In this paper we have presented a complete analysis  of operators up to (mass) dimension 6 describing spin 2 theories (e.g. weak field limit of theories linear and quadratic in the curvature), analyzing with some care the conditions for the theory to be (transverse) diffeomorphism invariant, scale invariant, conformal invariant and Weyl invariant. We have also identified a possible non-linear completion of those lagrangians.
\par
Conformality on shell is attained for any combination of the constants appearing in the dimension 4 and dimension 6 cases. The trace of the energy-momentum tensor is a total derivative, and besides the virial current for specific lagrangians is also the derivative of a two-index tensor, leading to improved forms of the corresponding energy-momentum tensors.
\par 
On the other hand, Weyl invariance instead does impose constraints on the coupling constants. Our main conclusion is to confirm \cite{Karananas,Farnsworth:2017tbz,Wu:2017epd,Budini} that Weyl invariance and conformal invariance are independent symmetries: not every Weyl invariant theory is conformal invariant in the weak field limit and conversely, not every conformal invariant theory is Weyl invariant in spite of the fact that it is always  invariant under global such Weyl transformations. To illustrate the first part of this statement, let us take for example the following WTDiff invariant theory
\be
\int d^4x\, \left(- \dfrac{1}{2 \kappa^2 } R[g^{-1/4}g_{\mu\nu}] + R^2[g^{-1/4}g_{\mu\nu}]\right)
\ee
where the precise form of these terms after permoning the transformation of the metric can be found in \eqref{linWTDiff} and \eqref{Weylquadratic}. The weak field expansion of this theory will contain, at quadratic order in the perturbation, dimension 4 operators and dimension 6 operators coming from the linear and quadratic (in curvature) pieces respectively. Theories combining operators of different dimension are not scale invariant, as pointed out in the example in \eqref{dim46}.  
\par
The analysis of dimension 5 and dimension 6 operators does not bring anything new with respect to diffeomorphism invariant theories, as expected. However, we have given expressions for the most general Lorentz and Weyl invariant lagrangians up to dimension 5 and dimension 6 operators, and we can clearly see that those theories contain an increasing number of arbitrary constants.
We have also discussed global Weyl invariance and it is clear that this symmetry is less restrictive than the local one. 
An analysis of the interaction terms has been done. It can be seen that potentials with diffeomorphism and weyl invariance can be constructive iteratively, for every orther of the perturbative expansion. 
\par
To end up, let us stress that the conditions that are argued to be neccessary for a ghost free non-local theory \cite{Biswas} are not compatible with the ones stemming from diffeomorphism invariance. 
\par
We finally point out that our results prove that {\em any} Lorentz invariant lagrangian for spin 2 particle up to quadratic order in the field is conformal invariant.

\section*{Acknowledgments}
 We have benefited from comments by Mario Herrero-Valea. This work has received funding from the Spanish Research Agency (Agencia Estatal de Investigacion) through the grant IFT Centro de Excelencia Severo Ochoa SEV-2016-0597, and the European Union's Horizon 2020 research and innovation programme under the Marie Sklodowska-Curie grants agreement No 674896 and No 690575. We also have been partially supported by FPA2016-78645-P(Spain). RSG is supported by the Spanish FPU Grant No FPU16/01595.

\newpage

\appendix
\section{Weak-field  limit of geometric scalars} \label{A}

We are interested in the expansion of the geometric invariants when we expand the metric around Minkowski spacetime
\be
g_{\m\n} = \eta_{\m\n} + \kappa h_{\m\n} 
\ee
If we take the limit of linear theories of gravity up to quadratic order in the fluctuations we have
\be
- \dfrac{1}{ \kappa^2 }\sqrt{g}\,R=\pd^\a\pd^\b h_{\a\b}-\Box h+h^{\a\b}\left({1\over 4}\Box \left(\eta_{\a\b}\eta_{\m\n}-\eta_{\a\m}\eta_{\b\n}\right)+{1\over 2} \eta_{\a\m}\pd_\b\pd_\n-{1\over 2}\eta_{\m\n} \pd_\a\pd_\b\right)\,h^{\m\n} + O(h^3)
\ee
When considering {\em TDiff} scalars one can also have terms of the type
\be
{\left(\nabla g\right)^2\over g^2}= \kappa^2 (\pd h)^2+ O(h^3)
\ee

The existence of this operator gives one extra freedom. To build the action which is {\em WTDiff } invariant we perform a Weyl transformation in the usual Einstein Hilbert action taking $\tilde{g}_{\m\n}= g^{-1/4} g_{\mu \nu}$ so that
\be
S_{WTDiff} =- \dfrac{1}{ \kappa^2 } \int d^4 x \, g^{1/4} \left(R + \dfrac{3}{32} \, \dfrac{(\nabla g)^2}{g^2}\right)
\label{linWTDiff}
\ee
Expanding it up to quadratic order in the fluctuations and writting it in terms of the four dimensional operators \eqref{4dimop} we get
\be
S_{WTDiff} = \int d^4 x \left( \mathcal{D}_1  +\mathcal{D}_2 +\dfrac 1 2  \mathcal{D}_3 + \dfrac{3}{8} \mathcal{D}_4 \right)
\label{WTDiff}
\ee

On the other hand, taking into account that
\be
R_{\m\n\a\b}={\kappa\over 2}\left(-\pd_\a\pd_\m h_{\n\b}+\pd_\a\pd_\n h_{\m\b}+\pd_\b \pd_\m h_{\a\n}-\pd_\n\pd_\b h_{\m\a}\right)+ O(h^2)
\ee
we learn that
\bea
&R_{\m\n\a\b}^2=\dfrac{\kappa^2}{ 4}\bigg\{4 {\cal O}_1-8 {\cal O}_3+4  {\cal O}_4\bigg\} + O(h^3)
\eea
For the Ricci tensor we have 
\be
R_{\n\b}={\kappa\over 2}\left(-\Box h_{\n\b}+\pd_\l\pd_\n h^\l_\b+\pd_\l \pd_\b h^\l_\n-\pd_\b\pd_\n h\right)+ O(h^2)
\ee
so that
\be
R_{\a\b}^2={\kappa^2\over 4}\left(2 {\cal O}_1-2 {\cal O}_2-2 {\cal O}_3+{\cal O}_4 +{\cal O}_5\right) + O(h^3)
\ee
Finally the expansion of the Ricci scalar reads
\be
R= \kappa \left( \pd^\a\pd^\b h_{\a\b}-\Box h  \right) + O(h^2)
\ee
and it follows that
\be
R^2=\kappa^2 \left( {\cal O}_1-2 {\cal O}_2+{\cal O}_5  \right)+ O(h^3)
\ee

A useful relationship is given by
\bea
&\a R_{\m\n\r\s}^2+\b R_{\m\n}^2+\g R^2= \kappa^2 \left[\left(\a+{\b\over 2}+\g\right)\,{\cal O}_1 - \left({\b\over 2}+2 \g\right)\,{\cal O}_2 -\left(2 \a +{\b\over 2}\right)\,{\cal O}_3+ \right.\nonumber\\
&\left. + \left(\a+{\b\over 4}\right)\,{\cal O}_4+\left({\b\over 4}+\g\right)\,{\cal O}_5 \right] + O(h^3)
\label{quadratic}
\eea
Using this it can easily be seen that the Euler density vanishes at this level of the expansion, whereas the Weyl squared tensor decomposes into
\bea
W_4=R_{\m\n\r\s}^2-2 R_{\m\n}^2+ {1\over 3}\,R^2={\kappa^2 \over 6}\bigg\{2\,{\cal O}_1+2\,{\cal O}_2-6\,{\cal O}_3+3\,{\cal O}_4-\,{\cal O}_5\bigg\} \nonumber \\
\eea

If we again consider quadratic theories which are {\em TDiff} invariant, one would have terms of the type must add
\be
{\left(\Box g\right)^2\over g^2}=(\Box h)^2
\ee
again, this yields in this case one extra freedom. \\

We can make the same analysis for the quadratic invariants but when considering actions that are {\em WTDiff} invariant. This can be achieved by making a Weyl transformation $\tilde{g}_{\m\n}=\Omega^2 g_{\m\n}$ on the usual quadratic action \eqref{quadratic} and then taking $\Omega^2 = g^{-1/n}$. For a general $\Omega$ we have 
\bea
&&\a \tilde{R}_{\m\n\r\s} \tilde{R}^{\m\n\r\s}+ \b  \tilde{R}_{\m\n} \tilde{R}^{\m\n} + \g \tilde{R}^2 = \Omega^{-4} \left(\a R_{\m\n\r\s} R^{\m\n\r\s}+ \b R_{\m\n}R^{\m\n} + \g R^2  \right) \nonumber \\
&& + \Omega^{-5} \left(-8\a -2(n-2)\b\right)R_{\m\n} \nabla^\m\nabla^\n \Omega + \Omega^{-6} \left(4\a + (3n-4)\b +4(n-1)^2\g \right) (\Box \Omega)^2 \nonumber \\
&&+ \Omega^{-6} \left(4(n-2)\a + (n-2)^2 \b\right)\nabla_\m\nabla_\n \Omega \nabla^\m\nabla^\n \Omega + \Omega^{-6}\left(-4\a -2(n-3) \b \right. \nonumber \\
&& \left. -2(n-1)(n-4)\g \right) R \nabla_\m \Omega \nabla^\m \Omega + \Omega^{-6} \left(16\a + 4 (n-2)\b\right) R_{\m\n} \nabla^\m\Omega \nabla^\n \Omega  \nonumber \\
&& \Omega^{-7} \left(8(n-3) \a+ 4(n^2-5n+5) \b + 4(n-1)^2(n-4) \g \right) \Box \Omega \nabla_\m \Omega \nabla^\m \Omega \nonumber \\
&& + \Omega^{-7} \left(-16(n-2) \a -4 (n-2)^2 \b \right) \nabla_\m\nabla_\n\Omega \nabla^\m \Omega \nabla^\n \Omega + \Omega^{-5} \left(-2\b -4 (n-1) \g \right) R \Box \Omega \nonumber \\
&& \Omega^{-8} \left(2n(n-1) \a + (n-1)(n^2-5n+8) \b + (n-1)^2 (n-4)^2 \g \right) (\nabla^\m \Omega \nabla_\m \Omega)^2
\label{Weylquadratic}
\eea
Using that $\Omega = g^{-1/2n}$ (in order to have {\em WTDiff}) and keeping dimension six operators with four derivatives and two metric fluctuations, we get for $n=4$
 \bea
 &&\a \tilde{R}_{\m\n\r\s} \tilde{R}^{\m\n\r\s}+ \b  \tilde{R}_{\m\n} \tilde{R}^{\m\n} + \g \tilde{R}^2  = \kappa^2 \left[ \left(\a+{\b\over 2}+\g\right)\,{\cal O}_1 +\dfrac{1}{2}(\a-\g)\,{\cal O}_2 \right.  \nonumber\\
 &&\left. -\left(2 \a +{\b\over 2}\right)\,{\cal O}_3+ \left(\a+{\b\over 4}\right)\,{\cal O}_4-\left(\dfrac{20\a+4\b -4\g}{32}\right)\,{\cal O}_5 \right]+O(h^3) \nonumber \\
 \eea
These are the most general theories that possess {\em LWTDiff}.

\section{Dimension 5 and dimension 6 operators with two-derivatives} \label{56}
This set of operators does not appear in the expansion of terms quadratic in the Riemann tensor, although they appear in the expansion of the Einstein-Hilbert lagrangian. 

For dimension 5, there are 14 independendent operators (up to total derivatives), that form a basis to expand the most general Lorentz invariant lagrangian containing such opeators
\begin{equation*}
\begin{aligned}[c]
&{\cal N}_1=  \, h^{\m\n} \pd_\n h_{\m}^\l \pd_{\l} h \nonumber\\
&{\cal N}_2=  \, h^{\m\n}  h_{\m}^\l \pd_{\l}\pd_\n h \nonumber\\
&{\cal N}_3=\,  h^{\m\n}  h \pd_{\m}\pd_\n h\nonumber\\
&{\cal N}_4=  \, h^{\m\n} \pd_\l h_{\m}^\l \pd_{\r} h^\r_\n\nonumber\\
&{\cal N}_5= \,  h^{\m\n}  h^{\r \l} \pd_{\r}\pd_\n h_{\m\l}\nonumber\\
&{\cal N}_6=   \, h^{\m\n}  h^{\r \l} \pd_{\r}\pd_\l h_{\m\n}\nonumber\\
&{\cal N}_7=   \, h^{\m\n}  h^{ \l}_\m \pd_{\r}\pd_\l h_{\n}^\r\nonumber\\
\end{aligned}
\qquad \qquad
\begin{aligned}[c]
&{\cal N}_8= \,  h^{\m\n}  h \pd_{\n}\pd_\l h_{\m}^\l \nonumber\\
&{\cal N}_9=  \, h^{\m\n}  h_{\m\n}  \pd_{\r}\pd_\l h^{\r\l} \\
&{\cal N}_{10}=   \, h^2 \pd_{\r}\pd_\l h^{\r\l}\nonumber\nonumber\\
&{\cal N}_{11}= \,  h^{\m\n}  h_{\m}^\l \Box h_{\n\l }\nonumber\\
&{\cal N}_{12}=   \, h^{\m\n}  h  \Box h_{\m \n}\nonumber\\
&{\cal N}_{13}=   \, h^{\m\n}     h_{\m \n}\Box h\nonumber\\
&{\cal N}_{14}=  \,  h^2 \Box h \\
\end{aligned}
\end{equation*}
For dimension 6, there are 38 independent operators (up to total derivatives), that form a basis to expand the most general Lorentz invariant lagrangian containing such opeators
\begin{equation*}
\begin{aligned}[c]
&{\cal K}_1= \,  h_{\a\b}h^\l_\n h_{\m\l}\pd^\m\pd^\n h^{\a\b}\nonumber\\
&{\cal K}_2=   \, h h^\l_\n\pd^\n h\pd^\m h_{\m\l}\nonumber\\
&{\cal K}_3=   \, h h^\l_\n\pd^\m h\pd^\n h_{\m\l}\nonumber\\
&{\cal K}_4= \, h_{\n\l}h^{\r\l}h_{\m\r}\pd^\m\pd^\n h\nonumber\\
&{\cal K}_5=  \, h h^\l_\n h_{\m\l}\pd^\m\pd^\n h\nonumber\\
&{\cal K}_6=  \, h h^{\l\a}h_{\l}^\b \Box h_{\a\b}\nonumber\\
&{\cal K}_7= \,  h^2 h_{\m\n}\pd^\m\pd^\n h\nonumber\\
&{\cal K}_8=  \, h_{\a\b}h^{\a\b}\pd_\m h\pd^\m h\nonumber\\
&{\cal K}_9= \,  h_{\a\b}h^{\l\b}\pd^\n h^\a_\n\pd^\m h_{\m\l}\nonumber\\
&{\cal K}_{10}=  \,  h h^{\r\l}\pd^\n h_{\n\l}\pd^\m h_{\m\r}\nonumber\\
&{\cal K}_{11}=  \, h^\a_\n h^{\l\b}\pd^\m h_{\m\l}\pd^\n h_{\a\b}\nonumber\\
&{\cal K}_{12}=  \, h_{\a\b}h^{\a\b}\pd^\n h^\l_\n\pd^\m h_{\m\l}\nonumber\\
&{\cal K}_{13}=  \,  h_{\a\b}h^{\a\b}\pd^\m h\pd^\n h_{\m\n}\nonumber\\
&{\cal K}_{14}=  \,  h^{\l\b}h^\a_\n h_{\m\l}\pd^\m\pd^\n h_{\a\b}\nonumber\\
&{\cal K}_{15}=  \,  h h_{\n\l}h_{\m\r}\pd^\m\pd^\n h^{\r\l}\nonumber\\
&{\cal K}_{16}=  \,  h_{\m\n}h^{\b\r}h^\a_\r\pd^\m\pd^\n h_{\a\b}\nonumber\\
&{\cal K}_{17}=  \,  h h_{\m\n} h^{\a\b}\pd^\m\pd^\n h_{\a\b}\nonumber\\
&{\cal K}_{18}= \,  h_{\a\b}h^{\l\b}h^\a_\n\pd^\m\pd^\n h_{\m\l}\nonumber\\
&{\cal K}_{19}=  \,  h h_{\n\l}h^{\r\l}\pd^\m\pd^\n h_{\m\r} \nonumber\\
\end{aligned}
\qquad \qquad
\begin{aligned}[c]
&{\cal K}_{20}=  \,  h_{\a\b}h^{\a\b}h^\l_\n\pd^\m\pd^\n h_{\m\l} \nonumber\\
&{\cal K}_{21}=   \, h^2h^\l_\n\pd^\m\pd^\n h_{\m\l} \nonumber\\
&{\cal K}_{22}=  \,  h_{\a\b}h^{\b\r}h^\a_\r\pd^\m\pd^\n h_{\m\n}\nonumber\\
&{\cal K}_{23}= \,  h h_{\a\b}h^{\a\b}\pd^\m\pd^\n h_{\m\n}\nonumber\\
&{\cal K}_{24}=  \,  h^3\pd^\m\pd^\n h_{\m\n}\nonumber\\
&{\cal K}_{25}= \,  h_{\a\l}h^{\l\r}h_{\r\s}\Box h^{\s\a} \nonumber\\
&{\cal K}_{26}= \,  h_{\m\n} h_{\a\b}h^{\a\b}\pd^\m\pd^\n h\nonumber\\
&{\cal K}_{27}=  \,  h_{\a\b}h^{\a\b}h_{\r\s}\Box h^{\r\s} \nonumber\\
&{\cal K}_{28}=   \, h^2 h_{\a\b}\Box h^{\a\b} \nonumber\\
&{\cal K}_{29}=   \, h_{\a\b}h^{\l\a}h_{\l}^\b \Box h \nonumber\\
&{\cal K}_{30}=  \,  h_{\a\b}h^{\a\b}h\Box h\nonumber\\
&{\cal K}_{31}=  \, h^3\Box h \nonumber\\
&{\cal K}_{32}=  \, h_{\m\l}h^{\l\b}\pd^\m h^\a_\n\pd^\n h_{\a\b}  \nonumber\\
&{\cal K}_{33}=   \, h h^{\r\l}\pd^\m h_{\n\l}\pd^\n h_{\m\r} \nonumber\\
&{\cal K}_{34}=  \,  h_{\a\b}h^{\l\b}\pd^\m h^\a_\n\pd^\n h_{\m\l} \nonumber\\
&{\cal K}_{35}= \,  h h_{\n\l}\pd^\m h^{\r\l}\pd^\n h_{\m\r} \nonumber\\
&{\cal K}_{36}=  \,  h_{\a\l}h^{\l\r}\pd_\m h_{\r\s}\pd^\m h^{\s\a}\nonumber\\
&{\cal K}_{37}= \,  h_{\a\b}h^{\a\b}\pd^\m h^\l_\n\pd^\n h_{\m\l} \nonumber\\
&{\cal K}_{38}=   \, h_{\a\b}h^{\a\b}\pd_\m h_{\r\s}\pd^\m h^{\r\s}
\end{aligned}
\end{equation*}
Finally, taking all the contributions mentioned in the text, the most general Weyl invariant lagrangian up to dimension 6 operators reads
\bea
\mathcal{L}_{W_{6D}}&=&  \mathcal{L}_{W_{5D}} + \kappa^2 \left\lbrace b_{1}\mathcal{K}_{1}+b_{2}\mathcal{K}_{2}+b_{3}\mathcal{K}_{3}+b_{4}\mathcal{K}_{4}+b_{5}\mathcal{K}_{5}+b_{6}\mathcal{K}_{6}+b_{7}\mathcal{K}_{7}+b_{8}\mathcal{K}_{8}+b_{9}\mathcal{K}_{9}+ \right. \nonumber\\
&&+b_{10}\mathcal{K}_{10}+b_{11}\mathcal{K}_{11}+\frac{1}{2}\left(-3a_{2}-b_{1}-b_{10}+4b_{2}-3b_{4}-4b_{5}\right)\mathcal{K}_{12}+\frac{1}{8}\left(2a_{1}-4a_{2}-8a_{3}-\right.\nonumber\\
&&\left.-4a_{9}-8b_{2}-8b_{3}+c_2\right)\mathcal{K}_{13}+b_{14}\mathcal{K}_{14}+\frac{1}{4}\left(-12a_{1}-3a_{5}-4b_{10}-2b_{14}-2b_{9}\right)\mathcal{K}_{15}+\nonumber\\
&&+\left(-\frac{3}{2}a_{1}+3a_{2}+12a_{3}-b_{1}+4b_{5}+4b_{6}+16b_{7}-\frac{3}{8}c_2\right)\mathcal{K}_{16}+\nonumber\\
&&+\left(-3a_{3}-2b_{5}-2b_{6}-8b_{7}\right)\mathcal{K}_{17}+\left(8a_{1}+4a_{2}+2a_{5}+4b_{10}+b_{11}-b_{14}-4b_{4}+2b_{9}\right)\mathcal{K}_{18}+\nonumber\\
&&+\frac{1}{4}\left(20a_{1}-16a_{2}+3a_{5}-8b_{1}+4b_{10}+2b_{14}-12b_{4}-32b_{5}+2b_{9}+c_2\right)\mathcal{K}_{19}+\nonumber\\
&&+\left(\frac{3}{2}a_{1}-3a_{2}-12a_{3}-a_{9}+b_{1}-4b_{5}-8b_{6}-16b_{7}+\frac{3}{8}c_2\right)\mathcal{K}_{20}+\nonumber\\
&&+\frac{1}{16}\left(-2a_{1}+4a_{2}+48a_{3}+4a_{9}+8b_{2}+8b_{3}+32b_{5}+32b_{6}+64b_{7}-c_2\right)\mathcal{K}_{21}+b_{22}\mathcal{K}_{22}+\nonumber\\
&&+\left(-\frac{3}{8}a_{1}+\frac{3}{4}a_{2}+3a_{3}-\frac{1}{2}a_{9}-\frac{1}{2}b_{1}-\frac{3}{4}b_{22}+b_{5}+2b_{6}+4b_{7}-\frac{3}{32}c_2\right)\mathcal{K}_{23}+\nonumber\\
&&+\frac{1}{192}\left(12a_{1}-24a_{2}-160a_{3}+16a_{9}+8b_{1}-16b_{2}+24b_{22}-16b_{3}-64b_{5}-96b_{6}-\right.\nonumber\\
&&\left.-192b_{7}+3c_2\right)\mathcal{K}_{24}+b_{25}\mathcal{K}_{25}+b_{26}\mathcal{K}_{26}+b_{27}\mathcal{K}_{27}+\frac{1}{256}\left(16a_{1}+48a_{11}-32a_{2}+\right.\nonumber\\
&&\left.+32a_{3}+16a_{9}+16b_{2}-64b_{26}-64b_{27}+16b_{3}+64b_{5}+64b_{6}+256b_{7}-128b_{8}- \right.\nonumber\\
&&\left.-10c_1+3c_2\right)\mathcal{K}_{28}+\frac{1}{96}\left(-12a_{1}+16a_{11}+24a_{2}+96a_{3}-8a_{1}-24b_{22}+32b_{26}+32b_{5}+ \right.\nonumber\\
&&\left.+32b_{6}+128b_{7}-3c_2\right)\mathcal{K}_{29}+\frac{1}{256}\left(32a_{1}+16a_{11}-64a_{2}-224a_{3}+32a_{9}+16b_{1}-16b_{2}+\right.\nonumber\\
&&\left.+48b_{22}-64b_{26}-64b_{27}-16b_{3}-64b_{5}-128b_{6}-256b_{7}+128b_{8}-2c_1+9c_2\right)\mathcal{K}_{30}+\nonumber\\
&&+\frac{1}{3072}\left(-104a_{1}-176a_{11}+208a_{2}+608a_{3}-128a_{9}-32b_{1}+16b_{2}-96b_{22}+192b_{26}+\right.\nonumber\\
&&\left.+192b_{27}+16b_{3}+64b_{5}+192b_{6}+128b_{8}+38c_1-21c_2\right)\mathcal{K}_{31}+\nonumber\\
&&+\left(12a_{1}+4a_{2}+3a_{5}+8b_{10}+b_{11}+2b_{9}\right)\mathcal{K}_{32}+\left(2a_{1}-6a_{2}-2b_{1}-2b_{10}-6b_{4}-8b_{5} \right. \nonumber\\
&&\left.+\frac{1}{4}c_2\right)\mathcal{K}_{33}+\left(20a_{1}-8a_{2}+3a_{5}-4b_{1}+4b_{10}+b_{11}-12b_{14}-16b_{5}+3b_{9}+\frac{1}{2}c_2\right)\mathcal{K}_{34}\nonumber\\
&&+\left(a_{1}+b_{10}\right)\mathcal{K}_{35}+\left(-\frac{3}{2}a_{1}+3a_{11}+3a_{2}+12a_{3}-b_{1}+3b_{25}+4b_{26}+4b_{5}+4b_{6}+16b_{7}- \right.\nonumber\\
&&\left.-\frac{3}{8}c_2\right)\mathcal{K}_{36}+\frac{1}{8}\left(12a_{2}+4b_{1}+4b_{10}+16b_{3}+12b_{4}+16b_{5}-c_2\right)\mathcal{K}_{37}+\nonumber\\
&&\left. +\frac{1}{64}\left(-8a_{1}+16a_{2}+32a_{3}+16a_{9}+16b_{2}+16b_{3}-128b_{8}+2c_1-3c_2\right)\mathcal{K}_{38} \right.\rbrace + \mathcal{L}_{6\partial \partial \partial \partial}^W \nonumber  \\
\eea


\end{document}